### Introduction

Unique possibilities of very cold neutrons (VCN) for structural studies are discussed last three decades [1]. VCN have wavelengths $\lambda \geq 10^1$ nm comparable with nanoparticles' size. VCN diffraction characterises their structure, and low–angle scattering reveals particles assembly. In molecular systems (polymers) the VCN show directly the structure factors since the formfactor of structural element (monomer unit) $F(\mathbf{q}) \to 1$ at low momentum transfer. VCN characterise nanostructures without any distortions due to an interference of waves scattered from atomic size structural elements. As a rule VCN Bragg scattering on crystals is forbidden since the distances between atomic planes $d < \lambda_{VCN}/2 \sim 10^1$ nm are too small as compared to VCN wavelength. Thermal and cold neutrons (TN, CN) mainly have $\lambda > 2d$. Their scattering on crystalline structures is masked by multiple Bragg diffraction. At last VCN create a 3D-image of scattering object whereas NT and CN give only 2D-image.

### Three-dimensional diffraction of VCN on nanostructures

CN scattered to low angles have momentum transfer $|\mathbf{q}|=|\mathbf{k}-\mathbf{k_o}| \ll k_o$ much smaller the initial momentum $k_o$. The **q**–vectors lie practically in plane, $\mathbf{q} \perp \mathbf{k_o}$ (Fig. 1). For a particle with cross section $d\sigma/d\Omega \sim |F(\mathbf{q})|^2$ this SANS gives 2D intensity distribution $I(q_X,q_Y) \sim |F(q_X, q_Y, q_Z=0)|^2$ proportional to the squared Fourier–image of particle in $(q_X, q_Y)$–plane [2]. Using VCN we find the 3D–image of nuclear scattering length density $\rho(\mathbf{R}) = \rho(X,Y,Z)$ in particle's volume $F(q_X, q_Y, q_Z) \sim \int \rho(X, Y, Z)\exp[i(q_X X + q_Y Y + q_Z Z)]dXdYdZ$. For CN we have the $(q_X, q_Y, q_Z=0)$–plane and formfactor averaged along Z–axis $F(q_X, q_Y, 0) \sim \int \rho(X,Y,Z)\exp[i(q_X X + q_Y Y)]dXdYdZ$. From CN–SANS one can rebuild only the scattering length density distribution averaged in Z–direction over all sample's layers. To recognise the 3D–structure, one needs to collect the 2D scattering pictures at all sample–beam mutual orientations, but it is not possible to rebuild the 3D–structure of an arbitrary object. This is real in some particular cases using structural models. If the shape (size) of object change (fluctuations, relaxation etc.), a consecutive observation of structure is not correct. At a given $\mathbf{k_o}$ VCN give spherical diffraction picture (Fig.1): $(q_Z + k_o)^2 + q_X^2 + q_Y^2 = k_o^2$. To overlap the desirable **q**-volume related to the sample, we must increase the $k_o = 2\pi/\lambda$ (decreasing $\lambda$) from internal ($k_{omin}=2\pi/\lambda_{max}$) to external ($k_{omax}=2\pi/\lambda_{min}$) shell by detector moving from the level $h=h_{min}$ to $h=h_{max}$ (Fig.1).

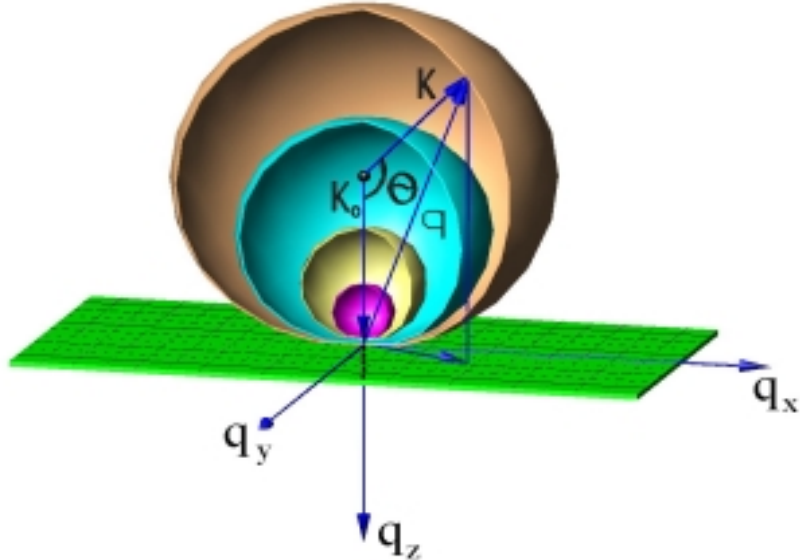

Fig. 1. VCN and CN scattering: 3D q–space for VCN and scattering plane for CN, θ is scattering angle

### Vertical VCN–diffractometer

To prepare collimated and monochromatic beams saving flux we propose to accelerate UCN in gravity field (Fig. 2) when neutron (mass m) drops down from the level h=0 to lower ones, h=1–25m. Neutron energy increases, $E_o \to E_h=E_o+mgh$, the line width remains constant, $\Delta E_h=\Delta E_o$. The Z-part of neutron velocity grows, $V_Z = (2gh)^{1/2}$, but the transversal projection, $V_\perp=(<V_X^2+V_Y^2>)^{1/2}=2(E_o/3m)^{1/2}$=const, that makes collimation, $V_{X,Y}/V_Z = (E_o/3mgh)^{1/2} = \Delta\theta_{X,Y}$. The collimation and line width are achieved without any loss of flux. It promises a progress of VCN in studies of soft matter and magnetic structures[3,4]. Polarised VCN can be produced by the reflection of UCN from magnetic field border and acceleration by gravity force. The VCN beams upgrade the resolution of 3D–polarimetry [5, 6] and Spin-Echo[7]. The VCN ($E_h\sim(1-10)\mu eV$, $\Delta E_h/E_h \sim 1-10$ %) are challenging instruments to probe condensed matter dynamics: by VCN inelastic scattering the sample's cross section $d\sigma/d\Omega d\omega \sim (k'/k_o)S(\omega,q)$ increases by two orders in magnitude due to large final neutron momentum $k'\gg k_o$.

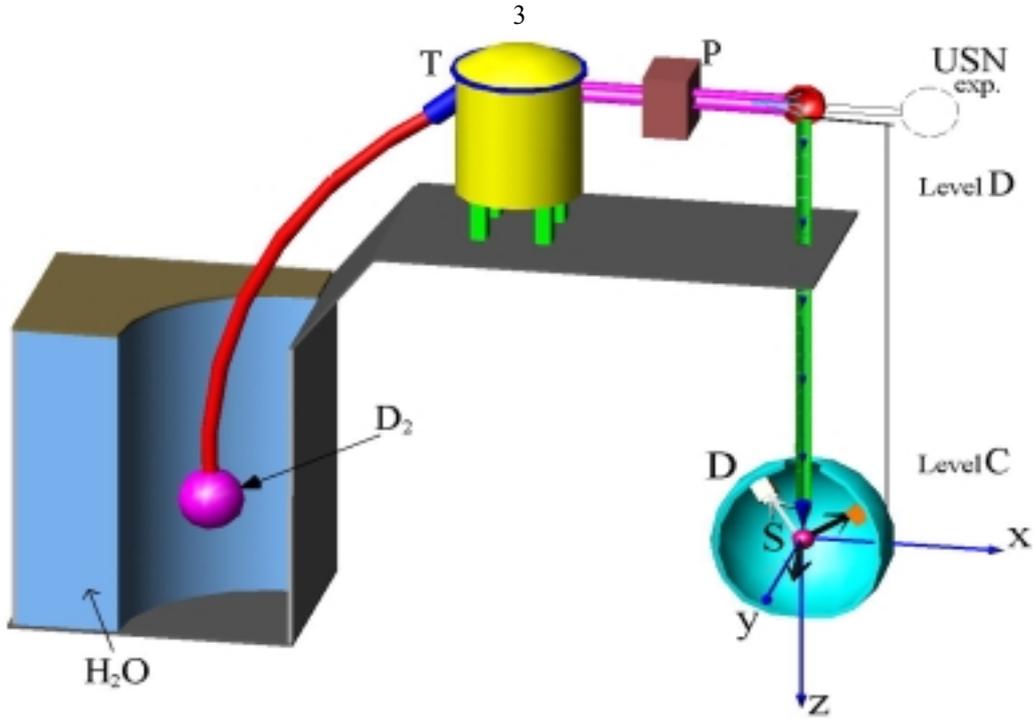

Fig.2. VCN–facility: guide (NG) conducts neutrons from the Cold Source ($D_2$) to turbine (T), UCN passed the polariser (P) and accumulated in the trap (D–level) fall down (C–level) taking acceleration, passing NG, collimator (path h), after scattering (sample S) VCN are detected (D). At D–level UCN are used in experiments

Parameters of VCN–facility for UCN with upper velocity modulus $V_b = 4$ m/s. In trap neutrons can fly up only to level $h_o = V_b^2/2g = 0{,}8$ m. Velocity components: $V_{X0,Y0,Z0} = V_o = V_b/\sqrt{3} = 2{,}3$ m/s. Divergence at level h: $\Delta\theta_{X,Y} = V_{X,Y}/V_Z = V_{X0}[V_{Z0}^2 + 2gh]^{-1/2} = [1 + 3h/h_o]^{-1/2}$. At $h_{min} = 1$m: vertical velocity $V_{ZMIN} = (2gh_{min})^{1/2} = 4{,}4$ m/s; $E_{min} = mg(h_o+h_{min}) = 0{,}18\mu eV$; $\lambda_{max} = 2\pi\hbar/(2mE_{min})^{1/2} = 67$nm; vertical momentum $k_{ozmin} = mV_{ZMIN}/\hbar = 0{,}07$nm$^{-1}$; divergence $\Delta\theta_{X,Y} = 0{,}46$ rad. $= 26{,}5°$; $\Delta\lambda/\lambda = (1/2)\Delta E/E = (1/2)h_o/(h_o + h) = 22{,}5$ %. At $h_{max} = 25$ m: vertical velocity $V_{ZMAX} = 22$ m/s; energy $E_{min} = mg(h_o+h_{min}) = 2{,}6\mu eV$; vertical momentum component $k_{ozmax} = 0{,}35$ nm$^{-1}$; $\lambda = 18$ nm; divergence $\Delta\theta_{X,Y} = 0{,}10$ rad. $= 5{,}9°$; line width $\Delta\lambda/\lambda = (1/2)\Delta E/E = (1/2)h_o/(h_o+h) = 1{,}6$ %. The spread $\Delta k_{ox,y,z} = mV_{X0,Y0,Z0}/\hbar = 0{,}037$ nm$^{-1}$ does not depend on h and define the resolution (Fig. 3).

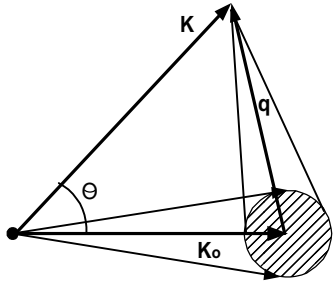

Fig. 3. VCN–scattering geometry by the initial spread of momentum

To improve the resolution we decrease the $\Delta k_{ox,y}$ keeping the necessary flux. For collimation $\Delta k_{ox,y} = 0{,}012$ nm$^{-1}$ better by factor 3 we get the flux 10 times smaller. At UCN density in trap $n \sim 10^3$ cm$^{-3}$ the initial neutron flux is equal to $f_n = (1/4)nV_b \sim 10^5$ cm$^{-2}$s$^{-1}$. The collimated beam with the flux $f_{nc} \sim 10^4$ cm$^{-2}$s$^{-1}$ is suitable scattering experiments. Detector resolution we keep higher than initial divergence limiting total resolution. Diffractometer has a set of removable NG between the trap and the collimator in front of detector. The collimator (diameter $d_C = 5$cm) forms the beam at the sample of the same diameter. Detector (solid angle $\Omega \sim 4\pi$) is composed of the elements ($d_D \sim 5$cm) fixed at the sphere (diameter $L_D \sim 1$ m). The element's resolution $\Delta\theta_D \sim 0{,}05$ rad. $= 2{,}9°$ is higher than initial divergence. The diffractometer (Fig. 1, 2) collects data in spherical q–volume $V_q = (4\pi/3)(k_{omax}^3 - k_{omin}^3)$. We can rotate the object to put its own **q**–volume to experimental one. Diffractometer quality is visible from the scattering simulation for a typical structure (Fig.4). Its scattering pictures were obtained for resolution: $\Delta q_{x,y} \approx \Delta k_{ox,y} = 0{,}012$ nm$^{-1}$; $\Delta q_z \approx \Delta k_{oz} = 0{,}037$ nm$^{-1}$. The available **q**–volume $V_q = (4\pi/3)(k_{omax}^3 - k_{omin}^3) = 0{,}18$ nm$^{-3}$ for $k_{omin} = 0{,}07$ nm$^{-1}$ and $k_{omax} = 0{,}35$ nm$^{-1}$ is built of the number of cells $N_q \sim 4\cdot10^4$, each volume $(\Delta q_x \cdot \Delta q_y \cdot \Delta q_z) \sim 5\cdot10^{-6}$ nm$^{-3}$.

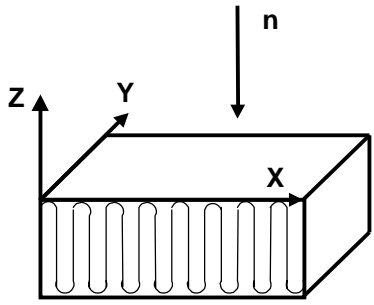

Fig. 4. Lamellar nanocrystal of polymer

The images of object (X,Y,Z-dimensions in proportion 1,5:0,75:1) are generated for positions (Fig.5–6). Level h =1m: neutron diffraction at small $k_{omin}$=0,07nm$^{-1}$ reflects mainly the size of object along the directions in 3D-space. To know the form and structure of object we scan its 3D q-space varying $k_o$ and measuring the diffraction at h =1–25 m. At a given $k_o$ an image shows a part of information. To get full data we overlap the q-space by a set of spheres, $k_{omin} \leq k_o \leq k_{omax}$. Very slow neutrons, scattered to an angle θ, have parabolic trajectories in gravity field and reach detector surface at the angular position $\theta_D < \theta$. The $\theta_D(\theta, t)$ depends on true scattering angle θ and neutron time of flight t: $R_D \cos\theta_D = V_Z \cdot t + g \cdot t^2/2$; $V_Z = V\cos\theta$. For scattered neutron with velocity components $V_X = V\sin\theta$, $V_Z = V\cos\theta$, the time t can be found from the equation $V^2 \cdot t^2 + gV\cos\theta \cdot t^3 + (g^2/4) \cdot t^4 = R_D^2$ where θ is defined by: $\cos\theta = (R_D/Vt)\cos\theta_D - gt/2V$. At small t≈R/V we have $\cos\theta \approx \cos\theta_D - gR_D/2V^2$ and the deviation ($\cos\theta_D - \cos\theta$) = (0,5-6,8) % at h=(25–1)m, detector radius $R_D$=50 cm. The images (Fig.5–6) display the pictures for true angles θ. Further compare the scattering at ideal and real resolution (Fig.7–10) to examine the diffractometer resolving abilities. The scattering functions $S_f(\mathbf{q}) = F^2(\mathbf{q})$ are plotted vs. variables $Q_{X,Y,Z} = q_{X,Y,Z}/q_{X,Y,ZMAX}$ in the q–ranges, $\Delta q_{x,y} \leq q_{x,y} \leq 0,35$ nm$^{-1}$, $\Delta q_z \leq q_z \leq 0,70$ nm$^{-1}$, covering nanoscopic scales $10 \leq 2\pi/q \leq 100$ nm.

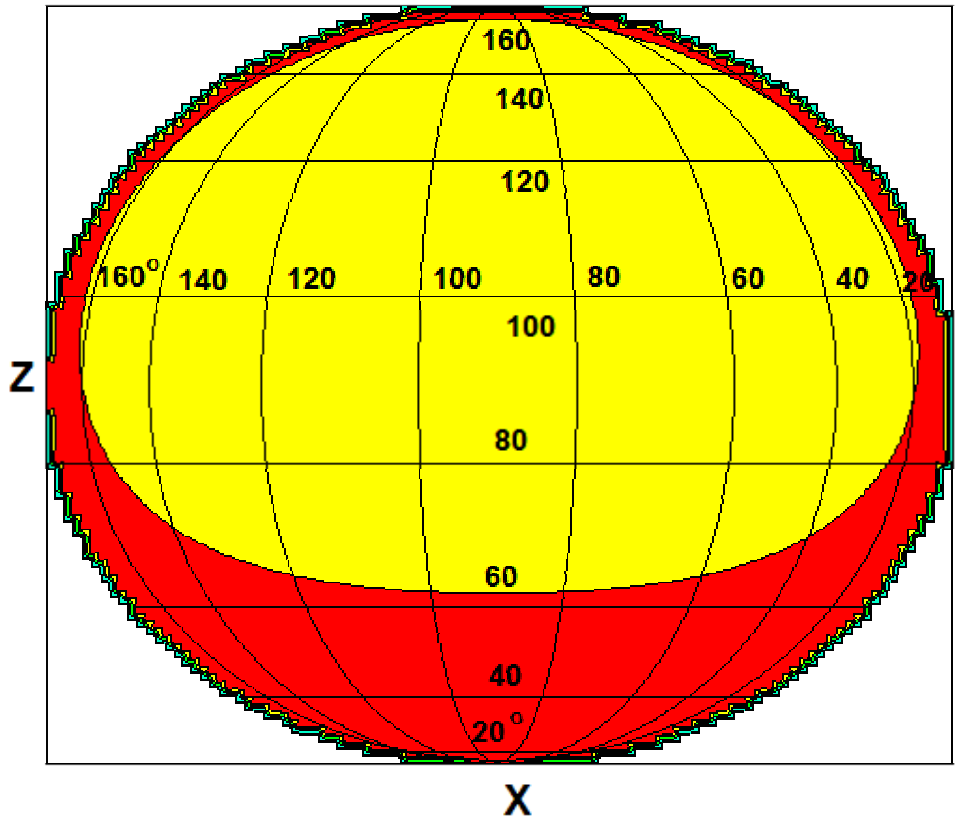

Fig. 5. Lamellar object image at detector (position $h_{min}$ = 1 m), surface projection to (X, Z)–plane is shown. VCN fall down along vertical Z–axis and illuminate detector's bottom by low-q scattering. Spherical angular coordinates: scattering angle θ = 20 – 160° (vertical); angle in (Y, X)–plane φ = 20 – 160°.

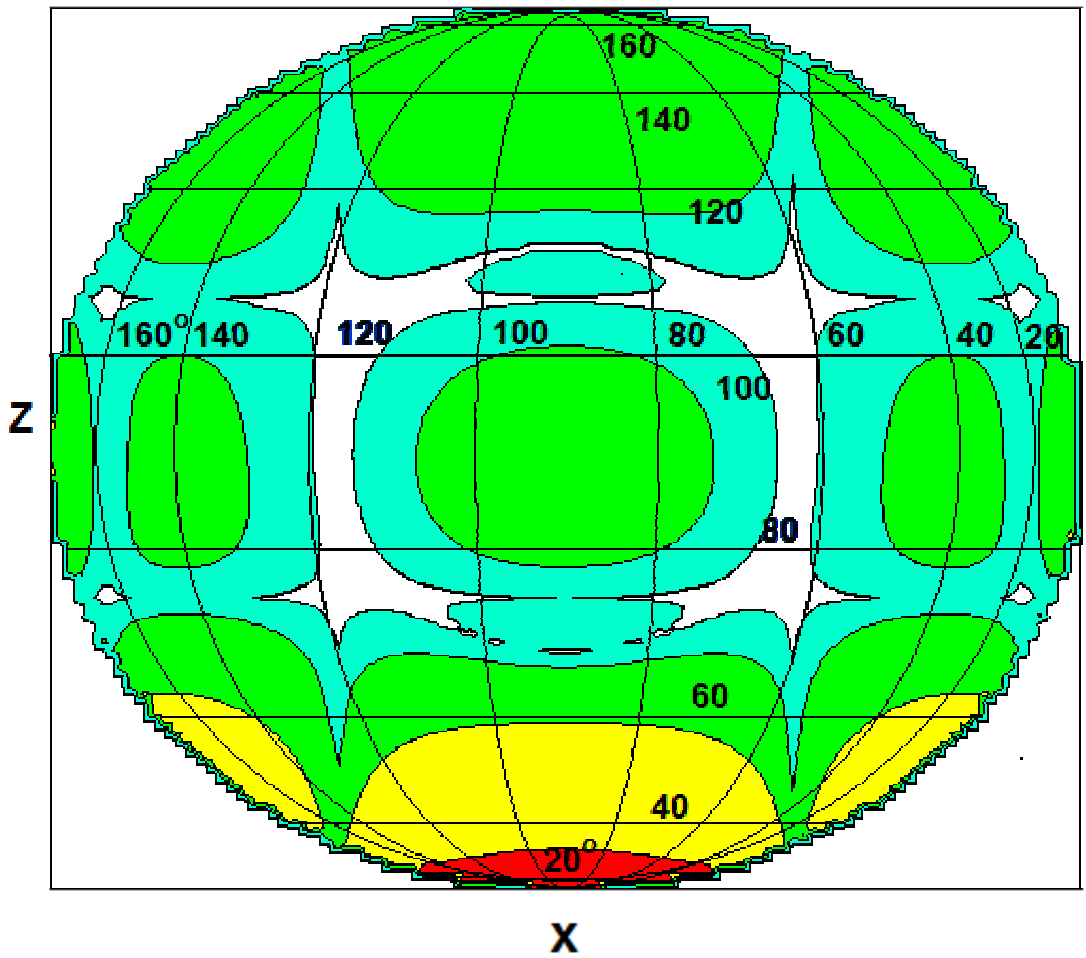

Fig. 6. Position $h_{max}$ = 25 m, image projection to (X, Z)–plane is displayed

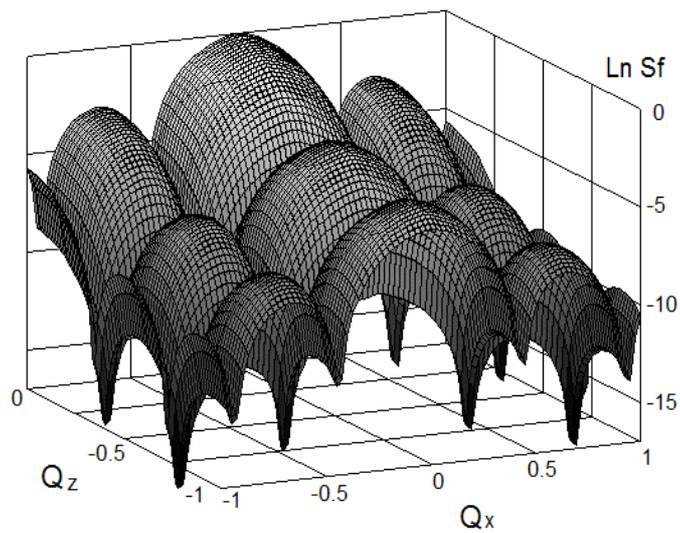

Fig. 7. Section ($Q_X$, $Q_Y = 0$, $Q_Z$) of scattering function $\ln S_f(\mathbf{Q})$ of lamellar object, ideal resolution

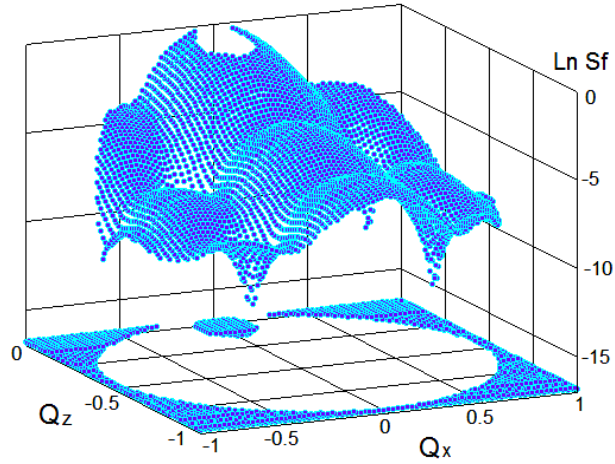

Fig. 8. Section ($Q_X$, $Q_Y = 0$, $Q_Z$) of scattering function $\ln S_f(\mathbf{Q})$ of lamellar object: real resolution

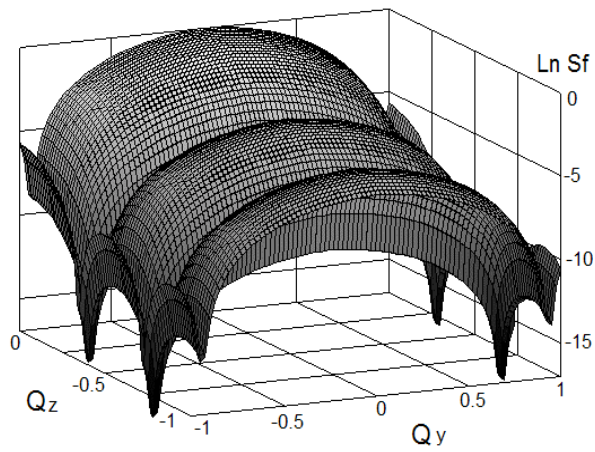

Fig. 9. Section ($Q_X = 0$, $Q_Y$, $Q_Z$) of scattering function $\ln S_f(\mathbf{Q})$ of lamellar object: ideal resolution

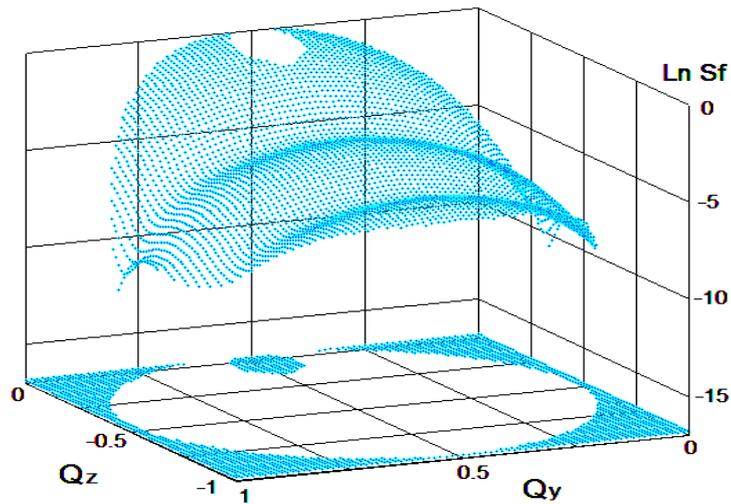

Fig. 10. Section ($Q_X = 0$, $Q_Y$, $Q_Z$) of scattering function $\ln S_f(\mathbf{Q})$ of lamellar object: real resolution

**Conclusions**

VCN gives a full information on object's structure transformed to 3D–Fourier image containing total atomic and molecular correlations. Vertical VCN–diffractometer, covering solid angle $\Omega \sim 4\pi$, integrates the experimental abilities of SANS and Reflectometry to study various nanostructures including horizontal liquid layers and interfaces, e.g. surfactants, biological and synthetic membranes, multicomponent structures with nanoparticles (fullerenes, nanotubes and derivatives), catalysts (chemistry, hydrogen energy).

**Acknowledgements**

The work was performed in terms of Russian Academy of Sciences Program "Neutron Studies of Structure and Fundamental Properties of Matter" and supported by Paul Scherrer Institute. Authors are grateful to RAS and PSI directorate.